\documentstyle[aps,twocolumn,psfig]{revtex}
\draft

\def\bef{\begin{figure}}
\def\eef{\end{figure}}
\def\bet{\begin{table}}
\def\eet{\end{table}}
\def\bea{\begin{eqnarray}}
\def\eea{\end{eqnarray}}
\def\ba{\begin{array}}
\def\ea{\end{array}}
\def\bi{\begin{itemize}}
\def\ei{\end{itemize}}
\def\ben{\begin{enumerate}}
\def\een{\end{enumerate}}

\begin{document}

\title{ Experimental Investigations of New Interaction \\
by Use of\\
Stationary High-accuracy Quartz Gravimeter}

\author{ Yu.A.BAUROV}
\address{ Central Research Institute of Machine Building\\
 141070, Moscow region, Kaliningrad, Pionerskaya, 4, Russia}

\author{ A.V. KOPAJEV}
\address{ Sternberg State Astronomical Institute\\
119899, Moscow, University prospect, 13, Russia}
\maketitle

\begin{abstract}
Smooth anomalous in time dependence recordings of a
high accuracy quartz gravimeter with a constant magnet attached
to it, were revealed. These anomalies of minute's duration have
amplitudes sometimes by more than an order of magnitude greater
than that of the Moon tide, and may not be explained with the
aid of current physical concepts.  The experimental procedure
was based on the hypothesis of a new interaction arising when
acting on physical vacuum by magnetic systems through their
vectorial potential.  The coordinates of physical space magnetic
anisotropy due to existence of the cosmological vector potential
${\bf A_g}$, a new basic vectorial constant, are determined. In particular,
the declination coordinate $\delta\approx 34^\circ $  of the vector
${\bf A_g}$ (second equatorial coordinate system) is determined for the first
time by experiment. The hypothesis considered was used for physical
justification of the results obtained.
\end{abstract}



\section{INTRODUCTION.}

In papers [1-5], the results of experimental
investigations of a new interaction discovered, are presented,
which arises when acting on the process of elementary particle
charge number formation by magnetic systems through their
vectorial potential. In classical and quantum field theory, the
potentials, as a rule, have no physical meaning by themselves
(cannot be measured), only their derivatives have it. In
particular, the vectorial potential of the electromagnetic field
is determined, in existing conventional theories, with an
accuracy of an arbitrary function gradient (gauge invariance),
which is closely related to the law of conservation of
elementary particle charge numbers.  It is shown [6-10] that if,
for example, electric charge $e(x)$ of an elementary particle is
varied over a set $\{x\}$, the potentials become single-valued.
Masses of elementary particles are found [9-10] to be
proportional to modulus of the cosmological vector potential
$\vert {\bf A_g}\vert \approx 1,95\times10^{11}$ CGSE units,
a new basic vectorial constant
appearing in one-dimensional discrete objects called byuons
\footnote{The former name is "one-dimensional discrete magnetic fluxes".},
from a finite aggregation of which, by a new paradigm \cite{11,12},
the entire surrounding world is formed, and specifically, the
three-dimensional space observed, the interior space of
elementary particles as well as their charge numbers are.

Now, if we direct the vector potential ${\bf A}$ of any magnetic system toward
the vector  ${\bf A_g}$ , we shall  meddle in the process of mass formation
of elementary particles, and, according to suggestion made in
Refs \cite{11,12}, a new force must appear to eject elementary
particles (and hence any substance) from a region with reduced
modulus of the cosmological vector potential  ${\bf A_g}$ , in the direction
of this vector. It is precisely this force that was detected in
experiments with strong resistive and superconducting magnets by
torsion and quartz piezoresonance balances.

In these experiments, the magnitude of the force was variable from $0.08g$
to $0.01g$ with magnetic fields $B$ from $1$ to $14 T$ in $40-53 mm$
diameter solenoid apertures and for various relative positions
of the torsion axis and test bodies (approximately $30g$ by mass)
with respect to the solenoid walls. One of two coordinates of
the vector ${\bf A_g}$ direction, the right ascension $\alpha \approx 270^\circ \pm 7^\circ$
(second equatorial system)
\footnote{Further, all astrophysical coordinates will be indicated only
in this coordinate system.}, was revealed, whereas the second
coordinate, the declination $\delta$, failed to be found experimentally
for lack of strong magnetic system with horizontal axis of
symmetry.

The vector ${\bf A_g}$ declination could be evaluated from
astrophysical observations. As the Sun has a strong magnetic
system with vector potential always directed (in some region)
toward  ${\bf A_g}$ , the Sun has  to move, under the influence of the new
force, in direction of ${\bf A_g}$ relative to the nearest stars \cite{11,12}.
This direction (the Sun's  apex) is known \cite{13} to have the
coordinates  $\alpha \approx 270^\circ, \delta \approx 30^\circ$. In Ref. \cite{5} the Sun's motion was
modelled with the aid of a superconducting magnetic system, but
only the angle $\alpha \approx 270^\circ - 300^\circ$ had been measured,
which was in qualitative accord with the Sun's apex coordinate $\alpha$. The
declination of ${\bf A_g}$ was estimated too by a revealed anisotropy of
solar flare distribution over the Sun's surface, which flares
are associated, as is known, with floating up of magnetic flux
tubes to the surface of the Sun under the action of buoyancy
force. This anisotropy turned out to be around $8\sigma$ where $\sigma$
corresponds to the uniform distribution of solar flares over the
Sun's surface, therewith $\alpha \approx 277^\circ, \delta \approx 38^\circ$ for it \cite{14}. The
anisotropy effect is attributed to action of the new force on a
magnetic flux tube during its floating up.

In this paper an attempt is made to determine the coordinates $\alpha$ and
$\delta$ of the vector ${\bf A_g}$ direction in terrestrial conditions with
the aid of a stationary high-accuracy quartz gravimeter.

\section{ THE DEFINITION OF  THE  PROBLEM AND  EXPERIMENTAL PROCEDURE.}

  As an experimental set-up, the tide gravimeter
developed in the Sternberg astronomical Institute of Moscow
University on base of a standard quartz gravimeter "Sodin"
(Canadian production) and described  in more detail in Refs.
\cite{15,16}, was used.  The experimental investigations were carried
out in a specialized gravimetric laboratory. The gravimeter was
placed in an underground room on a special foundation separated
from that of the building.  A schematic diagram of a quartz
sensitive system is shown in Fig.1. Its main component is a
quartz lever 1 ($2 cm$ in length with a platinum mass $m = 0.05g$)
suspended on torsion quartz fibres 2 and additionally off-loaded
by a vertical quartz spring 3. Such a  construction gives  to
the  device  high  sensitivity to changes in gravity up to $1 \mu Gal$
($10^{-8} ms^{-2} = 10^{-9}g, g$ is free fall acceleration) as well as
sufficient protection against microseismic disturbances.  The
optical recording system comprises a galogen lamp 4 fed from a
high-stabilized power source, light of which lamp enters into
the instrument through the objective lens 5 with the aid of
optical fibers. The quartz rod 6 welded to the level is a
cylindrical lens, it forms an image of the point light source,
which falls further on a photosensor rule 7 with a sensitivity
on the order of $1 \mu~m$. The digital signal from the photosensor
rule output enters further through a special interface to a
computer which executes preliminary processing and
averaging of data over the interval of $1 min$. The accuracy of
minute's data is $(0.5-1.2)\times 10^{-9}g$ depending on the microseismic
noise level varying considerably in the course of a day.  The
sensitive system of the instrument includes some additional
units (not shown in Fig.1) protecting it from thermal,
electrostatic, and atmospheric pressure disturbances.
Calibration of the instrument is carried out by the inclination
procedure with an accuracy of about $(0.5-1.0)\%$ which is quite
sufficient for the experiment being considered.  The amplitude
of constantly recorded by the device changes in gravity due to
Moon-Sun tides and corresponding Earth's deformation, is
$(50-200)\times 10^{-9}g$, which allows to evaluate magnitudes of possible
anomalous effects against the background of tides.  To measure
the new interaction by the Sodin gravimeter, a constant magnet
  ($60 mm$ in diameter, $15 mm$ in height, the field $B$ in the center
  of $0.3 T$) was attached to it in such a way that the vector
  potential lines of the magnet in the vicinity of a test
  platinum weight (see Fig.1) should be directed perpendicular
  to the Earth's surface (i.e. towards the vertical component of
  the vector ${\bf A_g}$ ), since it was assumed on base of astrophysical
  data (see above) that the declination $\delta$ for  ${\bf A_g}$  is around
  $30^\circ - 38^\circ$.

In Fig.2 the technical diagram of the experiment is  presented.
If the declination $\delta$ of the vector ${\bf A_g}$  is near $30^\circ$,
two nearly opposite positions of the gravimeter will  arise (for
the latitude of Moscow ($\approx 56^\circ$)) during Earth's daily rotation as
associated with the arrangement of the gravimeter sensitivity
axis with respect to the vector ${\bf A_g}$  (see Fig.2: positions A and
B).  In the position A, the new force ${\bf F}$, hypothetically collinear
with ${\bf A_g}$ , will be directed along the sensitivity axis of the
gravimeter, whereas in the position B it will be perpendicular
to this axis. If the values of $\delta$ measured should be much more
or less than $30^\circ$, the new force may manifest itself in the
position B, too.  To estimate the magnitude of the new force {\bf F},
acting on the test platinum weight of mass 0.05g, the following formula
was used \cite{1,2}:
$$ |{\bf F}| = 2 N \Phi m_{\nu_e} c_0^2 \frac{1}{A_g}
\frac{\partial \Delta A}{\partial x_1} \left( 1 - \frac{\Delta A}{A_g} \right),\eqno{(1)}$$

where $N$ is a number of stable particles (protons, neutrons,
electrons) in the test body;

$m_{\nu_e} c_0^2$ is the minimum residual potential
energy of four-contact interaction of byuons forming the
interior space of an elementary particle, which space is
identified \cite{11,12} with the rest mass of a pair "electron
neutrino-antineutrino"  ($\approx 33 eV$);

$\Phi = \frac{e^2}{h c_0}\frac{x_0}{c_0 t^*}$ is part of
energy $m_{\nu_e} c_0^2$ which can be acted upon by vector potential  of the
magnetic system \cite{4,12} ;

$e$ is the elementary electron charge;
$h$ and $c_0$ being the Planck's constant and light speed,
respectively;
$x_0 \approx 10^{-17} cm$; $ c_0 t^* \approx 10^{-13} cm$;

$\frac{\partial \Delta A}{\partial x_1} $ is
the derivative of variation of ${\bf A_g}$ modulus due to vector potential
of the constant magnet at the location point of the platinum
weight, over the spatial coordinate in direction of the detector
(the potential {\bf A} was calibrated so that its magnitude on the axis
of the constant magnet should be zero).

In our case, it was assumed that
$$\frac{\partial \Delta A}{\partial x_1} \approx
\frac{B_1 r_1 - B_2 r_2}{\Delta x} \approx 10~\hbox{\rm CGSE units}, $$
where $B_1$ and $B_2$ are, respectively, the
magnitudes of magnetic field at the location points of the
platinum weight and detector (point "O" on the quartz fibre of
the gravimeter (Fig1)); $r_1, r_2$ are distances from the axis of
the constant magnet to the platinum weight and point "Ž",
respectively; ${\Delta x}$ is the distance between the platinum
weigh and point "O".  The estimating calculations by the formula
(1) show that the constant magnet is capable to create $F \approx 10^{-10}N$.
  With the Moon tide amplitude equal, on the average, to
$10^{-7}g$, the gravitational force of Moon is around $10^{-10}N$, too.
Therefore, at first glance, the force $F$, created by a constant
magnet, may be measured with the aid of the gravimeter. But, as
is shown by experiments [1-5] and calculations of the Sun's
motion velocity in its apex direction under the action of the
force $F$ [11-12], the formula (1) gives by an order of magnitude
higher magnitude of the force $F$, because it does not take into
account nonlinearity of the (${\Delta A}$ - dependence of mass change and
nonlocal character of the phenomenon itself \cite{12} ($F$ depends on
detector coordinates (point. "O" in Fig.1) and test body).  In
Refs. \cite{4,12} the following expression for $F$ is given, which
accounts for nonlinearity of interaction and estimates its
nonlocality :
$$ {\bf F} = - 2 N \Phi m_{\nu_e} c_0^2 \lambda (\Delta A)
\frac{\partial\lambda(\Delta A)}{\partial\Delta A}
\frac{\partial \Delta A}{\partial x_1}\frac{\hbox{${\bf A_g}$}}{A_g}, \eqno{(2)}$$
where
$$\lambda (\Delta A) = \sum_{k=1}^{\infty} \lambda_k \exp \left\{
- \left[ \frac{\Delta A}{A_g} \frac{r}{\Delta x} \left(\frac{1}{\Phi}\right)
\right]^k \right\} {\Delta A}^k,$$
$\lambda_k$ are dimensional coefficients of the set; $r$ is the mean radius
between the test body and the point "O" (Fig.1) measured from
the axis of constant magnet.  The theoretical computations by
formula (2) agreed satisfactorily with experiments  \cite{4,12}.  It
can be shown that at $k=1$, the magnitude $F \sim \Delta A \frac{\partial \Delta A}{\partial x_1}$.  Therefore, in
connection with that there exists actually, in the vicinity of
the Earth, certain summary potential ${\bf A_{\Sigma}}$ equal to the sum of
${\bf A_g}$ and magnetic field potentials of Earth ($A_E \approx 10^8$ CGSE units), Sun
($A_{\odot} \approx 10^8$ CGSE units),  Galaxy ($A_{G} \approx 10^{11}$ CGSE units) etc., and
defined by some law yet unknown, and that fluctuations of these
huge potentials are possible, one might hope the force $F$ to be
measured by a gravimeter with a constant magnet.  Therewith $\Delta A$
should be created by some natural sources, and
$\frac{\partial \Delta A}{\partial x_1}$
should by a constant magnet attached to the gravimeter, since
$A_\Sigma$ from spatial sources varies at immensely long distances, with
an infinitesimal value of $\frac{\partial \Delta A}{\partial x_1}$
at a point of the gravimeter location.

\section{EXPERIMENTAL   RESULTS.}

In Fig.3 a typical recording from the
Sodin gravimeter with the constant magnet, obtained at the
period between Dec. 29, 95 and Jan. 5, 96 is shown. The major
deflections of the gravimeter, being repeated every 24 hours,
are associated with the Moon's pull of gravity. Denote an
average amplitude of Moon tide by $L$, an amplitude of accidental
events, recorded by gravimeter and corresponding to an increase
in pull, by  $k L^+$,  nd that corresponding to a decrease in pull,
by $k L^-$, where $k$ is a factor indicating the value of deflection in
terms of Moon tide amplitudes. The event recorded on Dec. 30,
1995, at 15.00 (Fig.3) correspond to a local stroke in Moscow,
for example, machinery in building (local earthquake); the event
of Jan. 01, 1996 (10.30) was caused by a planetary earthquake.
As is seen from Fig.3, both events are oscillatory in character.
The event of Jan. 2, 96 recorded at 10.36 (here and below the
local solar time is used), is uncharacteristic as of an
earthquake signal so of any other known disturbance having been
encountered earlier when operating with a Sodin gravimeter
(noises in power, electronics, transition jumps in quartz,
etc.). The time interval of this event was 2 min, amplitude
nearly $0.2 L^+$. In a time of 2 minutes a pull smooth increasing took
place, and the readings of the gravimeter returned to a normal
Moon tide curve.  Fig.4 shows the results of an uninterrupted
experiment from Feb. 24 to Mar. 22, 96. Three events were
documented:  on the 28th of February, at 10.05;  4th of March,
at  10.58; 18th of March, at 20.54. Two last of them had a huge
amplitude ($13.6 L^-$ and $15.2 L^-$, respectively) and
$\Delta t \approx 10 min$. A time
profile of the event on 18th of March, 1996, is shown in Fig.5
at a more large scale of time.  As is seen from this figure, in
a time of nearly 10 min a smooth moderation of the Earth's
gravitational pull on the platinum weight of gravimeter took
place, and then the gravimeter readings returned to the Moon
tide curve. Similar is the event on 4th of March. In 1994-96
years, the experiments with the Sodin gravimeter and magnet
attached to it were interrupted. From 13th of April, 1996 the
experiment in consideration goes uninterruptedly. Its procedure
was improved: close by the gravimeter considered another one was
located, with no magnet. The first one recorded an event with an
amplitude of $1.8 L^+$ and $\Delta t \approx 2 min$ on 19th of April, 1996, at 7.27,
whereas the magnetless gravimeter has not recorded this event.
The events recorded by the gravimeter with magnet are shown in
Fig.6, all numerated chronologically and asterisked on year and day circles.

\section{DISCUSSION.}

 The smooth minute's variations recorded by Sodin
gravimeter with magnet cannot be due to known noises in this
instrument as well as due to known external factors like
earthquakes. Therefore, these recordings are proposed by us to
be attributed to manifestation of the new interaction in
accordance with the experimental procedure above considered. It
should be noted once again that the event on 19th of April, 1996
(7.27) being recorded by the gravimeter with magnet, has not
been documented by the magnetless instrument.  As is seen from
Fig.6, eleven events of the twelve recorded locate in a sector
A, which confirms, in accord with above methodology, the value
presumed of the $\delta$-coordinate for the vector ${\bf A_g}$ equal to
$30^\circ-38^\circ$.
The only event 11 (Fig.6) with the greatest deflection amplitude
of the Moon tide curve ($\approx 15.2 L^-$ ), falls within the sector B. The
event 10 with an amplitude of about $13.6 L^-$ therewith was yet in
the sector A, i.e. in the course of the event 11, such a change
in vector ${\bf A_{\Sigma}}$ direction occurred that the gravimeter, when in
position B, recorded a force directed up from the Earth's
surface. Hence, the coordinate $\delta$ increased up to $\delta \approx 34^\circ$ which
magnitude may be taken as a coordinate for the vector ${\bf A_g}$  if it
assumed to be parallel with ${\bf A_{\Sigma}}$.  It is also seen from Fig.6 that
there was no events between the groups of events 3, 4, 6 and 1,
2, 5, 7, 9, 10. Absence of events in this angular sector may be
explained by that the derivative $\frac{\partial \Delta A}{\partial x_1}$ (1, 2) equals zero in
the region of the maximum decrease in the vector ${\bf A_g}$  magnitude
due to the potential ${\bf A}$, which decrease being associated with some
cosmic event changing its right ascension coordinate $\alpha$ (extremum
point), hence the force $F = 0$ precisely in direction of ${\bf A_g}$  and
increases drastically on both sides of this direction.
Therefore, the formula (2) gives only an approximate direction
for {\bf F} assuming  ${\bf F}\|{\bf A_g}$ .  Fig.6 shows clearly the angle
$\alpha$ for the vector ${\bf A_g}$  to be around $270^\circ$. This result corresponds completely
to all previous experimental researches [1-5] as to
astrophysical observations \cite{13,14}.  Based on data of IZMIRAN
institute for the Earth's magnetic surrounding, an investigation
was carried out by us to find a synchronisation of the events
recorded with such known phenomena as magnetic storms, solar
flares, crossing by the Earth the boundaries of Sun's magnetic
field sectorial structure, i.e. with those phenomena capable to
produce considerable changes in vector ${\bf A_g}$  magnitude. The
investigation has shown that in this time there was no solar
flares, and none of the events corresponded to intersecting the
Sun's magnetic field sectorial structure boundaries, where, in
the vicinity of the Earth, the highest electric currents may
flow.  The nearest events documented by us were offset by
approximately one day (15th of Dec., 1994) and two days (4th of
March, 1995) from the well ascertained polarity change time
point of the sectorial magnetic field. Only two events (15th of
Dec., 1994 and 19th of Apr., 1995) correspond, with an accuracy
of the nearest day, to weak magnetic substorms on Earth.
Therewith the event on 19th of April, 1994, coincides, to an
hour, with the origin of a magnetic substorm, which could not
cause a change of ${\bf A_g}$  magnitude more than $10^6$CGSE units. It
should be noted that the four events observed from 18.12.95 till
21.12.95, inclusive (see Fig.6), preceded a phenomenon of a
shock wave type in solar corona which was recorded by
radio-frequency emission.  Hence, the events of a minute's
duration, recorded by us, do not correspond to usual magnetic
phenomena near the Earth recorded by measuring magnitudes of
magnetic field $B$. We believe therefore to have discovered a
fundamentally new natural information channel predicted in Refs.
\cite{11,12,17,18} and associated with physical space structure
changes caused by ${\bf A_{\Sigma}}$  variations measured by a gravimeter as
manifestations of a new force. ${\bf A_{\Sigma}}$  may vary, for instance, due to
disturbances of the toroidal component of the Sun's magnetic
field from which $B = 0$ in Earth's orbit but the vector potential
exceeds $10^8$CGSE units, or due to some yet unknown events in
Galaxy and Universe. The experiments on the new force
investigation by gravimeter are being continued.  The authors
are grateful to L.S.Kuzmenkov for the discussion of the results
obtained, as well as to V.N.Ishkov for data presented on near Earth magnetic conditions, and to V.D.Jushkin for assistance in experimental work.

\newpage

Subscripts to figures of the article "Experimental
Investigations of New Interaction by Use of Stationary
High-accuracy Quartz Gravimeter" by Yu.A.Baurov and A.V.Kopajev:
\vskip10pt
Fig.1 Sensitive system of the Sodin quartz gravimeter:
 1   -	 beam, \\
2   -   horizontal quartz wires	 suspension system, \\
3   -   main spring, \\
4   -	 lamp, \\
5   -   object lens, \\
6   -   beam, \\
7   -	 CCD-SCALE,\\
 8   -   prism, \\
9, 10  -   thermocompensator, \\
11, 12, 13, 14 - micrometric compensation mechanism, \\
15  -  constant magnet.

\vskip10pt

Fig.2 Methodological diagram of the experiment:\\
 1 -  Earth's surface, \\
2 -  gravimeter, \\
3 -   sensitivity axis of  the gravimeter, \\
$\phi$ - Moscow latitude.
\vskip10pt

Fig.3 Readings of the gravimeter from Dec. 29, 1995 to Jan. 05,
	     1996, inclusive.\\
  $y$  -  the displacement of   platinum weight. \\
One division is equal to $0.1\mu$ and corresponds to $0.2\mu Gal$; \\
$x$  -  time (in minutes).
\vskip10pt

Fig.4 Readings of the gravimeter from the 24th of February to
	   the 22nd of March, 1996, inclusive.  \\
$y$  -  the   displacement of platinum weight. \\
One division is   equal to $0.1\mu$ and corresponds to $0.2\mu Gal$; \\
$x$	     -  time (in minutes).
\vskip10pt

Fig.6 Total~combination~of~events recorded by the gravimeter
	   with constant magnet: \\
\noindent
$1  -  15.12.94 (13.42) (0.9L^-,\Delta t \approx 10min);$ \\
$2  -  18.12.95 (14.02) (0.06L^+,\Delta t \approx 2min);$   \\
$3  -  19.12.95 (10.46) (0.07L^-,\Delta t \approx 2min);  $   \\
$4  -  21.12.95 (10.00) (0.05L^-,\Delta t \approx 2min);  $     \\
$5  -  21.12.95 (13.29) (0.1L^+,\Delta t \approx 2min);   $       \\
$6  -  02.01.96 (10.36) (0.2L^+,\Delta t \approx 2min);   $         \\
$7  -  08.01.96 (15.36) (0.8L^+, \Delta t \approx 2min $ (main peak), \\
total time $\approx 15min)$; \\
$8  -  09.01.96 (11.48) (0.7L^-, \Delta t \approx 2min $ (main peak), \\
total time $ \approx 10min); $ \\
$9  -  28.02.96 (10.05) (0.125L^-,\Delta t \approx 2min);$\\
$10 -  04.03.96 (10.58) (13.6L^-,\Delta t \approx 10min);$  \\
$11 -  18.03.96 (20.54) (15.2L^-,\Delta t \approx 10min);$    \\
$12 -  19.04.96 (07.27) (1.8L^+ ,\Delta t \approx 2min).$\\

\vskip10pt

\vspace*{10cm}

\newpage

\vspace*{10cm}
\bef
\protect\hbox{\psfig{file=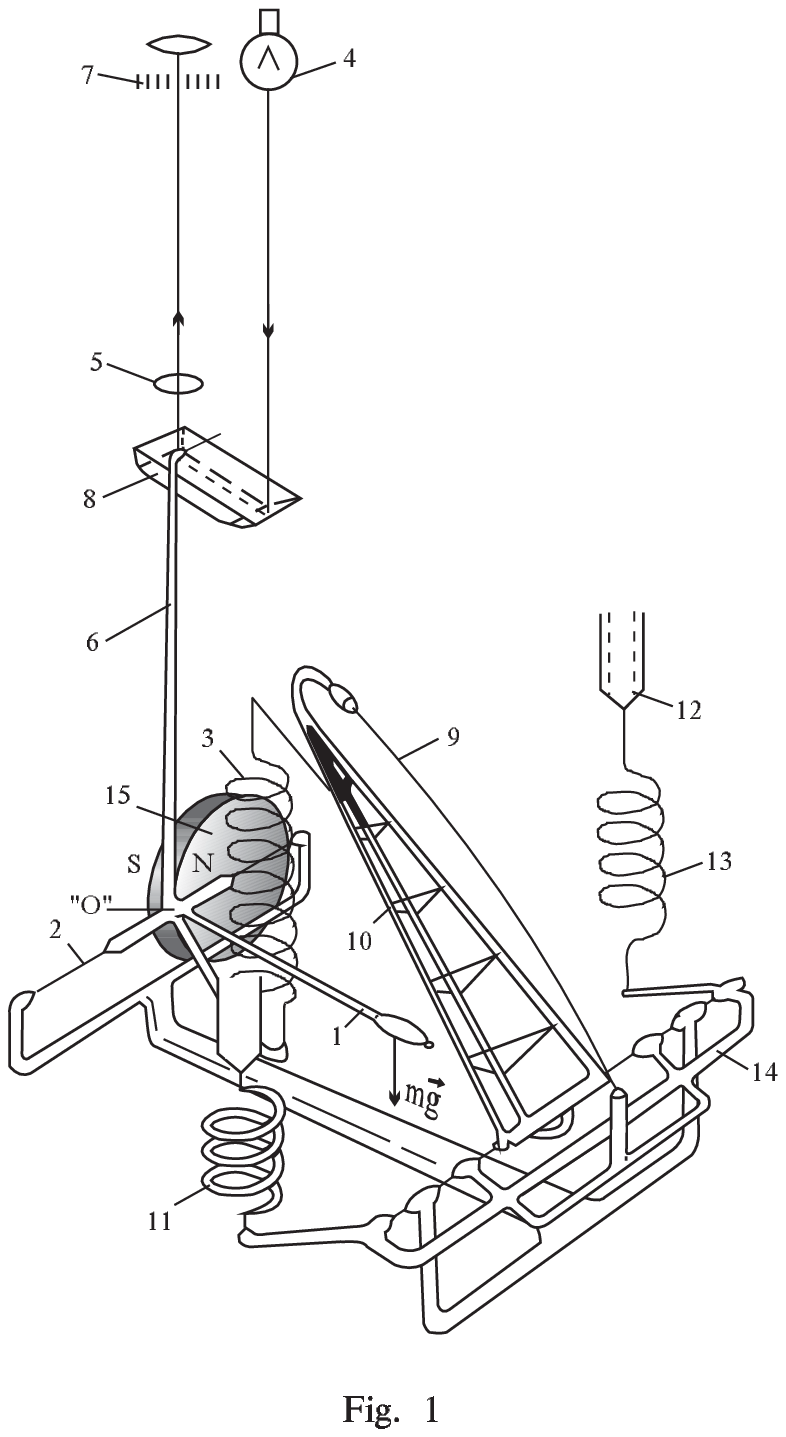}}
\label{1}
\eef
\draft
\vspace*{13cm}
\newpage
\bef
\protect\hbox{\psfig{file=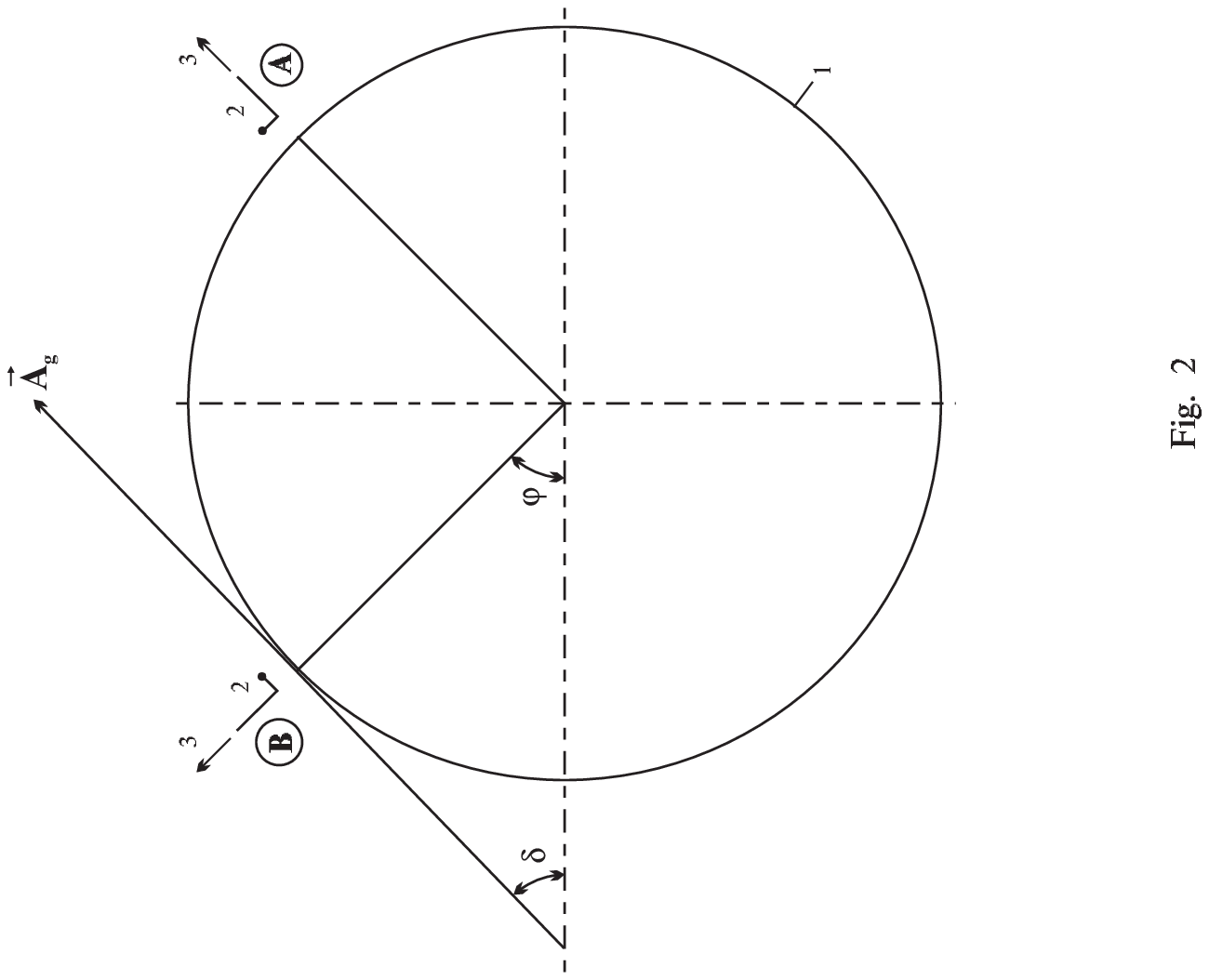,height=13.0cm,width=6.0cm,angle=90}}
\label{2}
\eef
\draft
\vspace*{13cm}
\newpage
\bef[t]
\protect\hbox{\psfig{file=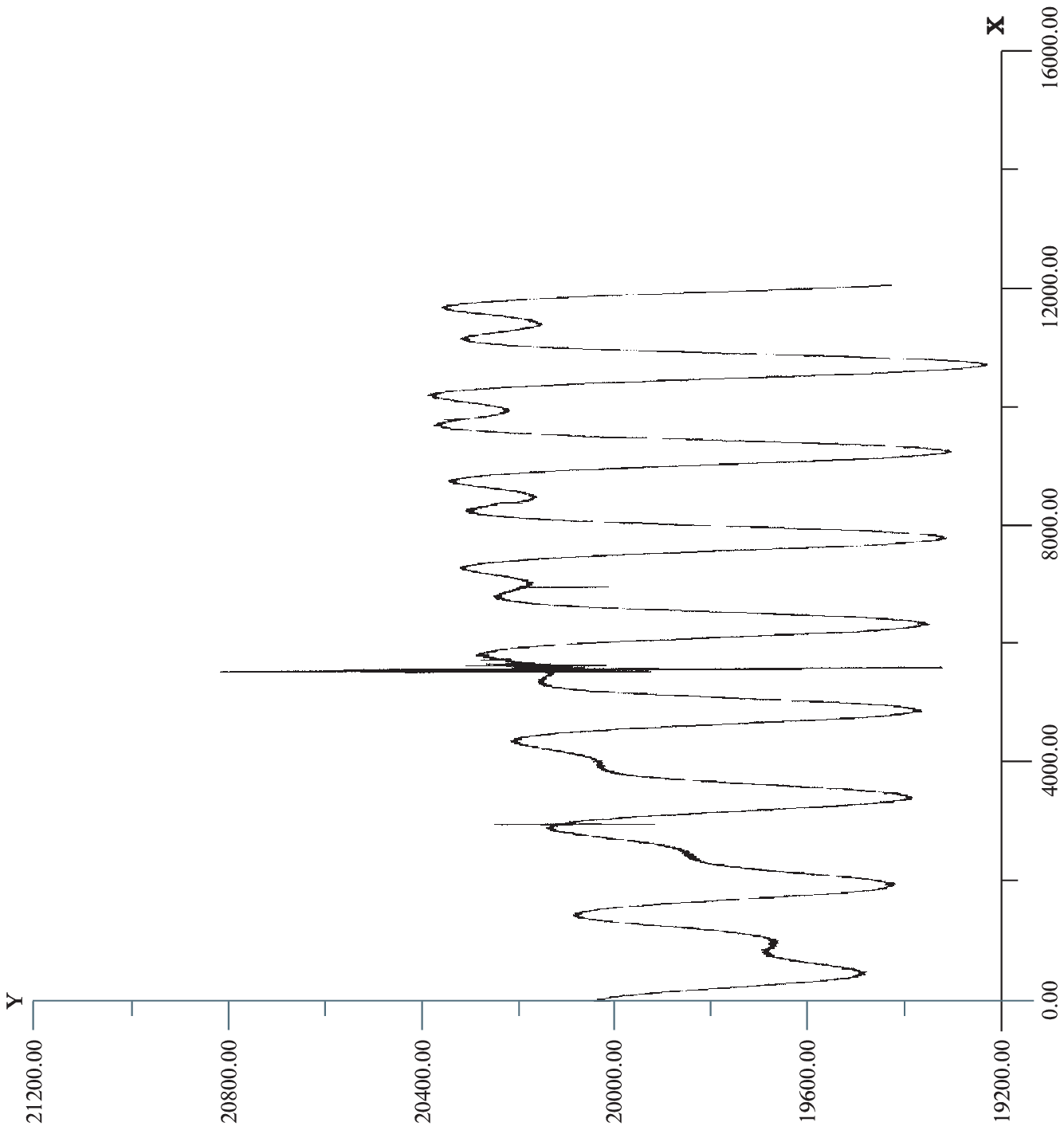,height=13.0cm,width=6.0cm,angle=90}}
\label{3}
\eef
\draft
\vspace*{13cm}

\newpage
\bef
\protect\hbox{\psfig{file=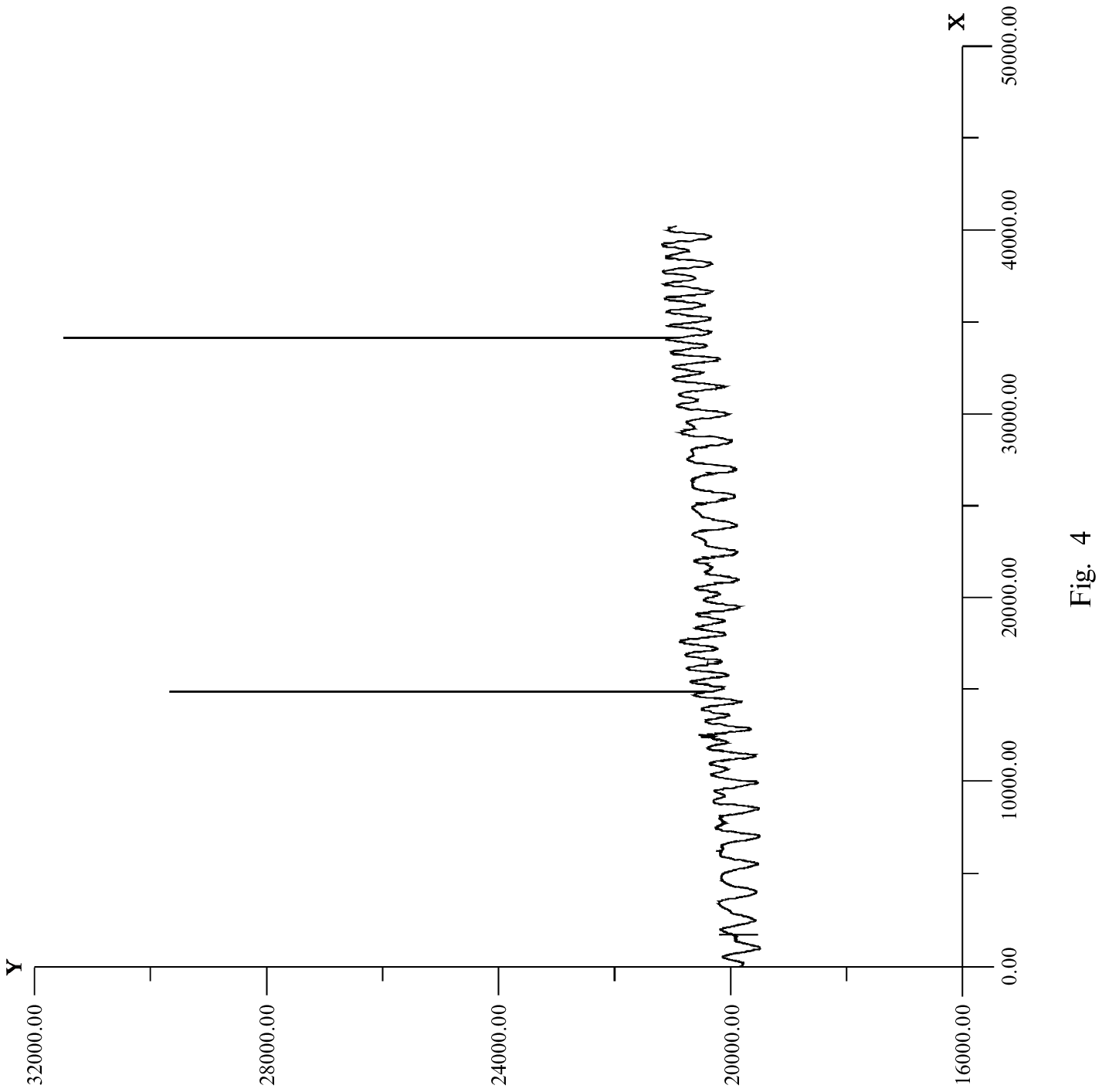,height=13.0cm,width=6.0cm,angle=90}}
\label{4}
\eef
\draft
\vspace*{13cm}

\pagebreak
\bef
\protect\hbox{\psfig{file=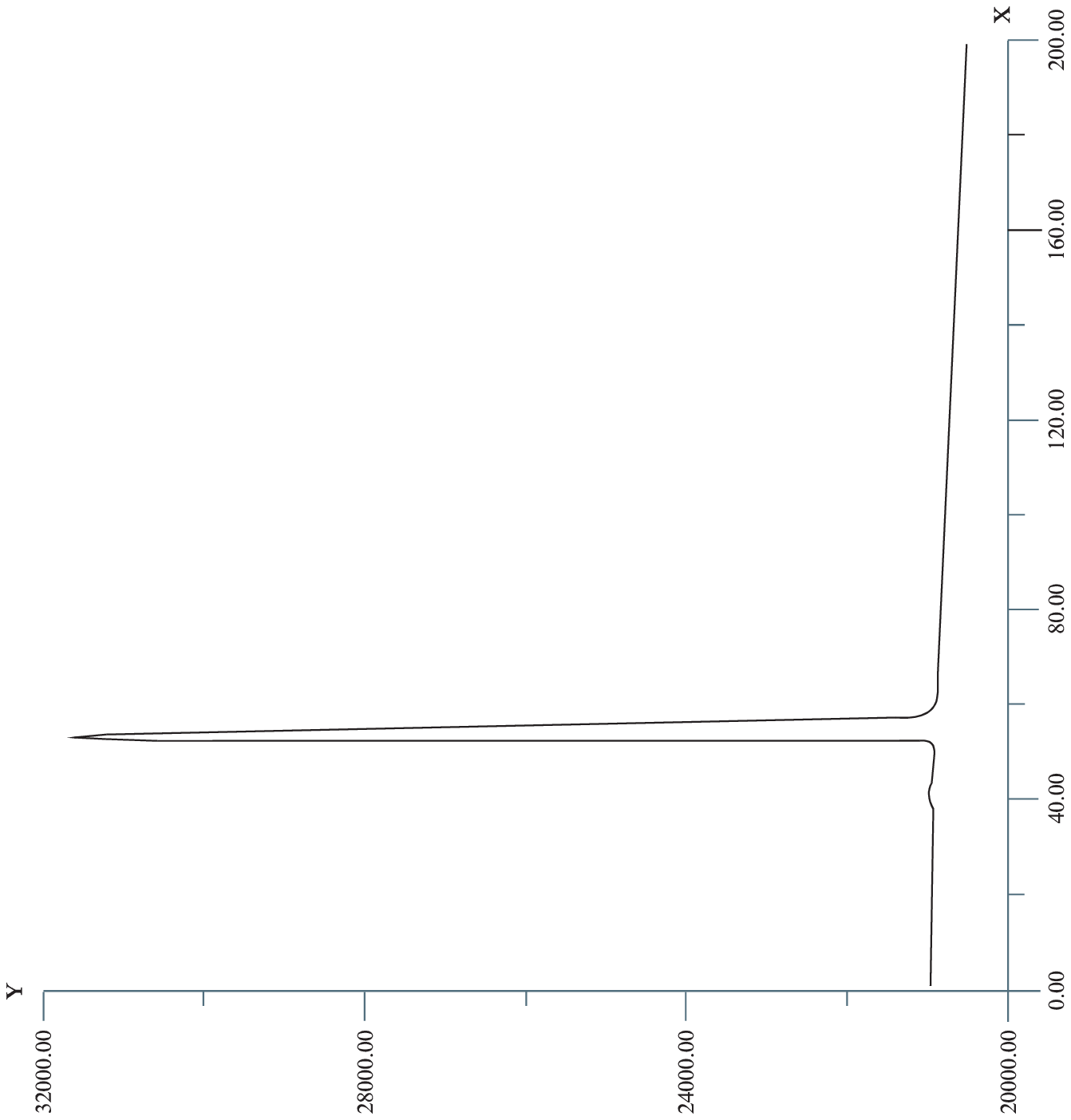,height=13.0cm,width=6.0cm,angle=90}}
\label{5}
\eef
\begin{draft}
\vspace*{13cm}
\newpage
\bef
\protect\hbox{\psfig{file=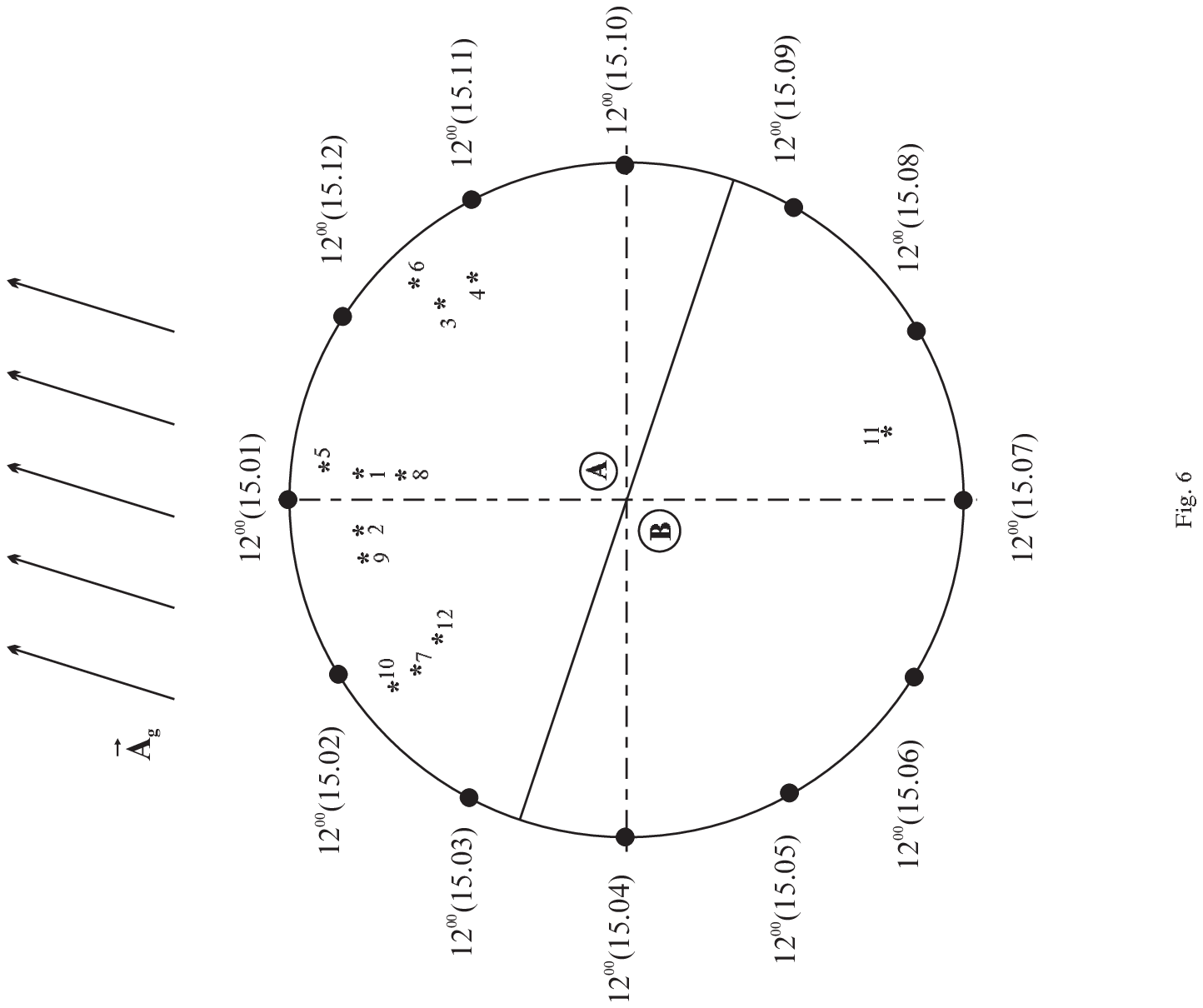,height=18.0cm,width=6.0cm,angle=90}}
\label{6}
\eef
\end{draft}

\end{document}